# Mapping the Sensitive Volume of an Ion-Counting Nanodosimeter

Reinhard Schulte[*a], Vladimir Bashkirov[a], Sergei Shchemelinin[b], Amos Breskin[b], Rachel Chechik[b], Guy Garty[c],
Andrew Wroe[d] and Bernd Grosswendt[e].

[a]*Dept. of Radiation Medicine, Loma Linda University Medical Center, Loma Linda, CA 92354*
[b]*Department of Particle Physics, Weizmann Institute of Science, Rehovot 76100, Israel*
[c]*Radiological Research Accelerator Facility, Columbia Univ., Irvington, NY 10533, USA*
[d]*Centre for Medical Radiation Physics, Wollongong Univ., Wollongong, NSW 2522, Australia*
[e]*Physikalisch-Technische Bundesanstalt (PTB), Braunschweig D-38116, Germany*

**Abstract**

We present two methods of independently mapping the dimensions of the sensitive volume in an ion-counting nanodosimeter. The first method is based on a calculational approach simulating the extraction of ions from the sensitive volume, and the second method on probing the sensitive volume with 250 MeV protons. Sensitive-volume maps obtained with both methods are compared and systematic errors inherent in both methods are quantified.

**Submitted to *JINST*, Jan. 16  2006**

---

[*] Corresponding author: Ph. (909) 558-4243, Fax (909) 558-4083,  *rschulte@dominion.llumc.edu*

## 1. Introduction

Ion-counting nanodosimetry measures the number of single ions deposited by an incident ionizing particle or photon within a gas volume simulating a nanometric volume of condensed matter. In a recent publication we described the performance of an ion-counting nanodosimeter (ND), in which the radiation-induced individual ionizations of charged-particle tracks are counted in a millimeter-size, wall-less volume of dilute gas [1]. The use of mbar-range gas-pressure permits expanding the nanometric ionization track to millimeter-scale dimensions. This ND has potential applications in diverse fields that require knowledge of the mean number of ionizations and its fluctuations within nanometric volumes of condensed matter. Ion-counting NDs described in [1] have been installed on research and medical accelerator beam lines to investigate particle-induced ion-cluster statistics over a broad energy range. An important feature of the ND is that it comprises a completely wall-less sensitive volume (SV), tunable by the internal electric field configuration. The small SV is embedded in a much larger gas volume, from which radiation-induced ions are registered. For applications simulating a specific target volume, e.g., a DNA segment, it is important to map out the exact shape and size of the wall-less sensitive SV.

A schematic cross section of the ND is shown in Fig. 1. Ions are generated stochastically by traversing charged particles within the ND chamber filled with a working gas. The primary particles are tracked by a silicon tracking telescope (STT). Under the applied electric drift field $E_1$, the ions drift towards the cathode located at the bottom plate of the chamber, which contains a small (millimeter-sized) aperture at its center. During the drift process, the ions undergo collisions with gas molecules, which introduces a random component (diffusion) into the ion motion. A much higher electric field $E_2$, established below the aperture plate, accelerates them to an energy of 8 keV towards the ion counter (a vacuum electron multiplier), where they are individually detected and counted. The ion registration efficiency of a given ion is equal to the product of the probability that it drifts through the aperture (named *ion extraction efficiency*) and the probability of ion detection at the ion counter. The former depends on the starting point of the ion drift, and the latter has been measured to be close to 100% at 8 keV for the ion counter used in our ND [2]. In the following, we assume that the ion registration efficiency equals the ion extraction efficiency.

Figure 1

In the present work, the "SV map" is defined as the spatial distribution of the ion extraction efficiency within the ND gas volume. The detailed shape of the SV map is determined primarily by the diameter of the aperture and the electric field configuration within the gas volume; to a lesser extent it depends on the gas and on the transport parameters of the ions, such as drift velocity and diffusion.

Under typical ND operating conditions, the ND chamber is filled with 1.33 mbar propane; 1mm in the gas phase is equivalent to about 3 nm in the condensed phase (water, DNA), based on the gas-to-water density ratio. With an aperture of 1 mm diameter, the SV transverse dimension is in the 1.5-mm range and the ion extraction efficiency falls rapidly to zero outside this range (from 80% to 20% over a range 1-2 mm) [1]. Mapping of a wall-less SV with a spatial resolution of 0.1 mm dictated by this rapid lateral fall-off of the ion extraction efficiency is challenging. We solved this task using two independent methods: a *calculational* method based on solving ion-transport equations with measured ion-transport parameters, and a *scanning* method based on experimentally probing the SV with high-energy protons, individually and precisely localized with the STT.

## 2. The calculational method

In this approach, previously measured ion diffusion parameters [3,4] were used to derive a map of the SV. The extraction efficiency of ions originating from a given point in the ND chamber was calculated by simulating the trajectories of a large number of ions deposited at this point and counting the fraction of ions arriving within the limits of the ND aperture. The ion motion is described by a drift along the electric field direction, combined with a lateral diffusion according to a Gaussian distribution [5]. The electric field in the gas volume, which is defined by the potentials applied to all electrodes including those below the aperture plate (not shown in Fig. 1), was calculated using the SIMION code[1]. Thereby, the penetration of the electric field $E_2$ through the aperture was taken into account. The ion motion in the 3D-varying field was calculated by summing the simulated motion within a sequence of segments, sufficiently small to consider the electric field in each segment as homogeneous. In each segment, the drift along the field was combined with a random lateral displacement $x_i$ due to transverse diffusion, which was sampled from a Gaussian distribution with the *rms* displacement:

$$\langle x_i \rangle^2 = \frac{2\Delta L_i}{E_i} \frac{D_i}{K_i} \qquad (1)$$

where $\Delta L_i$ is the length of the *i*-th segment, $E_i$ the local electric field in the segment, $D_i$ the local transverse diffusion coefficient, and $K_i$ the local ion mobility. The ratio $D_i/K_i$ is a function of the reduced field strength $E_i/p_i$, where $p_i$ is the local gas pressure. For our calculations, the $D/K$ values for hydrocarbon ions in propane were required, which were not readily available from the previously published literature. Therefore, we used the values obtained from our own measurements described in [3,4]. There, we measured the lateral spread of gas ions produced by α-particles in propane and drifting in the parent gas under a homogeneous electric field for different values of $E/p$. The measured values $D/K$ as a function of reduced field strength are shown in Fig. 2. An example for simulated ion trajectories in propane is shown in Fig. 3.

Figure 2

Ion extraction efficiency distributions were calculated for a grid of points spaced 0.2 mm in lateral direction and 7 mm in vertical direction. Efficiency values at intermediate points were obtained by linear interpolation. For each grid point, a total of 5000 trajectories were calculated. The SV map was determined for two different electric field conditions, reproducing typical experimental conditions as described in detail in section 3.2.

Figure 3

## 3. The scanning method

### *3.1 Principle*

This experimental method is based on the measurement of the spatial distribution of the integrated ion extraction efficiency using individual energetic protons passing through the gaseous SV, perpendicular to and at a precisely known distance from its central axis. The ND mounted on the research beam line of the proton accelerator at Loma Linda University Medical Center (LLUMC) is equipped with a high-resolution STT, which allows reconstruction of the trajectories of the particles traversing the ND chamber within 0.1 mm [6,7]. Successful application of this approach requires that the probe particles (1) are minimally scattered in the tracking detectors placed in front and behind the gas volume, (2) have a small ionization probability to avoid multiple primary ionizations within the SV, and (3) contribute a small number of ions due to secondary (delta) electrons compared to those from primary ionizations.

---

[1] http://www.simion.com/

To meet these requirements, we used 250-MeV protons. Multiple Coulomb scattering in the silicon layers (400 μm) of the STT is not significant at this energy. For the SV size of about 1.5 mm diameter in the present ND, the average mass length of a track segment within the non-zero efficiency region is about 0.3 μg/cm$^2$. On the other hand, based on an ionization cross section of 4.654 × 10$^{-18}$ cm$^2$ [1], the mean-free-path mass length of 250 MeV protons between two successive primary ionizations in propane is about 16μg/cm$^2$. Thus, the likelihood that more than one ion is generated by the probe particle is small, and most observed events will have either none or only one registered ion. The contribution of ions generated by delta electrons depends on their energy and origin. When electrons with energies in the 100 eV – 1 keV range are produced by the probe particle within the SV, the probability of additional ionizations within the SV is negligible; these electrons, due to their long ranges (~0.30 - 4 μg/cm$^2$ in propane), typically escape the small SV region without inducing further ionizations. For energies below 100 eV, the electrons have a short range but their ionization probability is small due to the rapid increase of the *W*-value (the mean energy required per ionization) with decreasing energy [8]. There is, however, a non-negligible chance that energetic delta electrons produced at some distance from the SV will reach it and deposit a secondary ion not caused by the primary particle. Such ions, which introduce a systematic error in the measured ion extraction efficiency, can be suppressed in two ways: first by applying a position-dependent time cut on the ion arrival time [1], and second using a background-subtraction procedure, as described below.

*3.2 Experimental setup and conditions*

The ion-counting ND used for the scanning experiments, was described in detail in our previous work [1]. The STT, which was added later, comprises two Si strip detector (SSD) modules, placed in front and behind the ND, each consisting of two Si planes with strips oriented in horizontal and vertical direction respectively [6,7]. The inner Si planes are located at 13.5 cm from the SV axis on either side of the SV. The detectors have a pitch of 194 μm, and outer dimensions of 6.4 cm by 6.4 cm. The spatial resolution of the probe-particle track reconstruction is better than 0.1 mm.

For SV scanning data acquisition, a broad beam of 250 MeV protons was collimated with three pairs of lead bricks, forming a vertical slit of about 1 cm width, which was centered on the SV. Thereby, data collection from protons crossing the gas volume far from the central axis of the SV was avoided. The data acquisition system of the ND acquired data on an event-by-event basis: for each event a trigger from a probe particle started a drift-time interval of 100 μs, during which the ions drifting from their origin in the SV to the ion counter were registered and time-tagged. This was followed by a read-out window of 15 μs duration, during which the probe-particle coordinate and ion drift-time data were transferred to the data acquisition PC.

SV scanning data were acquired under two ND operating conditions. The first set of measurements was performed with a stationary drift field $E_1$ of 60 V/cm across the gas volume. A second data set was obtained with a pulsed-field extraction, in which a background extraction field $E_1$ of 20 V/cm was superimposed with a 200 μs field pulse of 40 V/cm following each particle trigger with a 5-μs delay. The purpose of the pulsed-field operational mode was to sweep out all the ionization electrons from the chamber volume under the low field (the 5 μs delay is largely sufficient with the current chamber size) before the ion extraction started under the high field. This pulsed-mode operation avoids multiplication of the electrons in the gas, inducing unwanted secondary ions, as demonstrated previously [1]. In addition, the thickness of the aperture plate was different in each experimental condition (0.3 mm in the first condition versus 0.1 mm in the second). The plate thickness affects the extraction efficiency via the electric field $E_2$ penetrating through the aperture and acting as a focusing lens for the ions just above the aperture. The thinner aperture plate was introduced to extend the high-efficiency region of the SV in the longitudinal direction.

The acquired raw data consisted of 16-bit binary words encoding the following information for each event: the number of strips recording a hit for each Si-detector plane, the number of probe-particle triggers within an event, a time tag of the first trigger in the event, the number of ions registered in the ND, and time tags for each registered ion. Each experimental run contained of the order of 50 million events (3.5 GB of data). After rejecting events containing pile-ups (i.e., events with two or more triggers), events with missing track information, events with more than two neighboring strips hit on an SSD module, and tracks inconsistent with the primary proton beam direction (scattered primary particles), about 35-40 million events were available for further analysis.

For each event accepted for analysis, the proton trajectory was reconstructed based on the coordinate measurements provided by the four SSD planes, measuring the horizontal ($x$) and vertical ($y$) particle coordinates upstream and downstream the SV. The trajectories were sorted according to their position into $y$-binning intervals of 1 mm and $x$-binning intervals of 0.1 mm; typically, there were about 5000 protons per bin. Further, the number of ions associated with each proton (typically zero or one) and their arrival time were registered. As expected, in a scatter plot of proton $y$-coordinate versus ion drift time the points are scattered around a line, marking the linear relationship between the average ion drift-time and its deposition $y$-coordinate (Fig. 4). To suppress ions generated by energetic delta electrons formed at some distance from the SV, an ion drift-time cut was performed by excluding ions with drift times outside the time interval marked by the diverging lines in Fig. 4.

Figure 4

## *3.3 SV Map Reconstruction*

In our coordinate scheme (Fig. 1), the probe particles move along the *z* direction with known lateral (*x*) and vertical (*y*) coordinates. Assuming that the registered ions originate only from primary ionizations along the probe-particle track, the mean number of ions, $\mu(x,y)$, recorded per proton is related to the integrated (along *z*) ion extraction efficiency $\Sigma(x,y)$ as

$$\Sigma(x,y) \equiv \int_{-L/2}^{L/2} \varepsilon(x,y,z)dz = \lambda\mu(x,y) \qquad (2)$$

where $\lambda$ is the mean-free-path length with respect to ionization in the low-pressure propane gas, and $\varepsilon(x,y,z)$ is the ion extraction efficiency at a given location. The integration boundaries correspond to the longitudinal distance between the SSD modules and the SV axis. Based on an ionization cross section of $4.654 \times 10^{-18}$ cm$^2$ [1], we assumed $\lambda = 6.468$ cm at 1.33 mbar propane.

Fig. 5 shows, as an example for a recorded data set, the mean number of registered ions per particle as a function of the lateral distance from the SV axis for a vertical height of $y = 14.5$ cm above the aperture plane. The graph demonstrates that the data were approximately Gaussian-distributed with a constant additive background. The background can be attributed to ions induced by delta electrons that were not excluded by the ion drift-time cut procedure. For each available *y* value, these data were thus fitted to a Gaussian function with an additive background:

$$\mu(x,y) = a_1(y)\exp\left[-a_2(y)(x-a_3(y))^2\right] + a_4(y) \qquad (3)$$

where the model parameters $a_1$ – $a_4$ were evaluated by a nonlinear least-squares fitting procedure[2]. The lateral coordinate of the Gaussian peak was found to be constant to within +/- 0.1 mm over the range of $y$ values, and its location was redefined as the central axis of the SV ($x = 0$). Furthermore, the first $y$ value where the mean value exceeded the background level was redefined as the position of the aperture ($y = 0$). The background-corrected model function $\mu_{bgc}(x,y) = a_1(y)\exp\left[-a_2(y)(x-a_3(y))^2\right]$ was multiplied by the mean free path length to obtain an estimate for the integrated ion extraction efficiency $\Sigma(x,y)$, according to equation (2).

Figure 5

Due to the cylindrical symmetry of the setup with respect to the SV axis, the efficiency distribution $\varepsilon(x,y,z)$ was assumed to be a radially symmetric function $\varepsilon(r, y)$, where $r = \sqrt{x^2 + z^2}$ is the radial distance from the SV axis. In this case, the integrated ion extraction efficiency represents the Abel integral transform of the efficiency distribution. The extraction efficiency distribution $\varepsilon(r,y)$ was reconstructed from its discrete projection values $\Sigma(x,y) = \Sigma(ia, jb)$ ($i, j$ = 1, 2, 3..., $a$ = 0.1 mm, $b$ = 1 mm) using a standard reconstruction algorithm for the Abel transform [9].

For direct comparison with the maps derived by the calculational approach (section 2), the scanned-map values were binned into 7-mm intervals in vertical direction and values for intermediate $y$-coordinates were derived by linear interpolation.

## *3.4 Monte Carlo Simulations*

In order to estimate the systematic error related to the contribution of ions from delta electrons that could not be fully suppressed by the above-mentioned ion drift time cuts and background subtraction, Monte Carlo (MC) simulations were performed with a dedicated proton track-structure simulation code developed for the ND [1]. The code was used to simulate the data measured with the scanning method, i.e., the mean number of registered ions per particle as a function of the lateral and vertical coordinates of the primary proton, for an *a priori* known SV map, for which the SV map obtained with the calculational approach under stationary drift field conditions was used. The integrated ion extraction efficiency, $\Sigma(x,y)$, derived from the MC simulation was then compared with that obtained by integrating the SV map. As the MC simulations included delta electrons producing ions at some distance from the primary-particle trajectory, the ion arrival time-cut procedure was included in the simulation by counting only ions with an arrival time falling within an interval of $\pm 2\sigma$ around the mean arrival time for the given proton-to-aperture $y$-coordinate. The standard deviation $\sigma$ and the mean of the arrival time were obtained from the experimental drift-time data shown in Fig. 4. Simulated data sets were generated for proton tracks at lateral distances $x$ from the SV axis ranging from 0 to 3mm with 0.2-mm intervals and at vertical distances $y$ ranging from 0.5 to 49.5 mm with 7-mm intervals. A total number of $10^7$ proton tracks were simulated for each track location and the mean number of registered ions per proton was determined for each location. As in the experimental procedure, the mean background number of ions for proton tracks beyond a 2-mm $x$ distance from the principal axis was subtracted. The integrated ion extraction efficiency, $\Sigma(x,y)$, was calculated using equation (2). This was done with and without ion arrival time cuts and background corrections.

---

[2] Minerr function of Mathcad 11, Mathsoft Engineering & Education, Inc., Cambridge, MA, USA.

Fig. 6 shows the results of the MC-simulation tests. The simulated integrated ion extraction efficiency is plotted versus the lateral distance $x$ from the SV axis for three different vertical $y$-coordinates and compared with the numerically integrated SV map used for the simulation.

These graphs demonstrate that without any correction the magnitude of the systematic error introduced by ions originating from delta electrons ranges from 20-30%. This error can be effectively suppressed to about 2-10% by applying the correction procedures. The correction is most effective at short distances from the aperture.

Figure

## 4. Comparison of SV maps derived with both methods

Fig. 7 shows a side-by-side comparison and the distribution of differences of the calculated and scanned SV maps obtained for the first experimental condition: a stationary electric drift field of 60 V/cm and an aperture plate thickness of 0.3 mm. The efficiency maps have a flame-like form, indicating that the extraction of ions is increasingly controlled by diffusion at larger distances from the aperture. Aside from a similar general shape, the two maps show some differences: the calculated map has a broader base than the scanned one and the region of high efficiency (>75%) does not extend as far above the aperture as in the scanned map. One should also note that the ion extraction efficiency of the scanned map exceeds 100% (up to 126%) near the aperture. This means, *de facto*, that more ions are registered than expected based on the assumed ionization cross section, which can be explained in part by gas multiplication of electrons accelerated within the drift field [1] and in part by a residual contribution from delta electrons.

Figure 7

Fig. 8 shows the maps and their differences obtained for the second experimental condition: pulsed drift field and an aperture plate thickness of 0.1 mm. In comparison to the maps of Fig. 7, the region of high efficiency extends farther above the aperture, which is expected since the thinner aperture plate results in better focusing by the penetrating field $E_2$ (Fig.3). Note that the scanned map obtained under the pulsed electric field condition stays near 100% in the high-efficiency region (maximum of 109%). This supports the assumption that for a stationary high drift field of the first experimental condition electron multiplication took place in the gas, leading to the excess of ions observed under these conditions; this effect was, in fact, strongly suppressed under pulsed-field conditions. Nevertheless, there is a residual, systematic excess of the ion extraction efficiency in the scanning method, most likely due to a residual contribution from delta electrons.

Figure 8

## 5. Discussion and Conclusions

Two conceptually different methods for mapping the SV ion extraction efficiency in an ion-counting ND were investigated under two different experimental conditions. The scanning method provides a direct measurement of the integrated ion registration efficiency but requires about 40 hours of accelerator beam-time to achieve a statistical accuracy of the ion extraction efficiency of about ± 5%. The calculational method is naturally more rapid and economic; it is, therefore, more practical for testing and optimizing the electric field and geometrical parameters. Nonetheless it is necessary to compare the results of the two methods under well defined conditions, which was one aim of this work.

The SV maps obtained with the two methods were of similar shape but showed distinct differences (Figs. 7 and 8), which are probably related to different systematic errors affecting each method. The largest quantitative differences between the scanned map measured under pulsed-field conditions and the calculated map were up to +18% in the central region of the SV and -23% in a small region at the base of the SV (see Fig. 8c). The scanning method yielded ion extraction efficiencies larger than 100% (up to 126%) in a static field mode, which can be attributed to secondary ions induced by electron multiplication within the ND gas volume, under the high static drift field [1]. This was significantly reduced (maximum efficiency of 109%) after implementing the pulsed-field ion extraction. Another systematic error of the scanning method, also leading to an overestimation of the ion extraction efficiency, is caused by ions induced by delta electrons that are falsely assumed to originate from primary ionizations. Since it was difficult to estimate this contribution from first principles, it was simulated with a known SV map and the integrated efficiencies derived from the simulated data were compared to those obtained by numerical integration of the SV map. It was shown that the excess of efficiency introduced by ions originating from delta electrons can be reduced to significantly less than 10%, in particular close to the aperture, by using appropriate ion arrival-time cuts and a background correction.

Additional errors in both methods could originate from the uncertainties in the parameters needed for calculating the ion extraction efficiency from either simulated or measured data. The accuracy of the calculational method is determined by the accuracy of the electric field simulation and that of the ion transport parameters. Inaccuracy of the field simulation may explain, in part, the observed difference in the detailed shape of the maps obtained with both methods. Ion transport parameters were not available in the literature, and, therefore, we used data from our own measurements. The accuracy of the $D/K$ parameter was estimated to be of the order of ±10% (Fig. 2); additional measurements may further improve the accuracy of this parameter. The accuracy of the scanning method depends on the accuracy of the mean-free-path length for ionizations, which was estimated to be about ±5%[3].

One of the goals of our work has been to measure the distribution of ion cluster sizes in equivalent DNA volumes of 10 – 50 base pairs of length (4 – 17 nm in condensed phase, 1 – 6 mm in the ND gas phase) and to use these distributions to predict the yields of single and clustered DNA damages in plasmids with an appropriate biophysical model [10]. An important question related to this goal is how accurately the SV map represents a model of a DNA molecule surrounded by water in its "efficiency" to convert radiation-induced ionizations to DNA damage both by direct ionization and indirectly via diffusing free radicals. With the configuration of the SV ion extraction efficiency revealed by both methods (Figure 8), it is reasonable to assume that we are approximately simulating the high efficiency of direct DNA ionizations to produce damage within a cylinder of about 2.3 nm diameter (0.8 mm in the ND gas phase), whereas the rapidly falling efficiency of diffusing water radicals and hydrated electrons to cause damage with distance from the DNA is represented by the lateral fall-off of the ion extraction efficiency in the SV. For example, the diffusion distance of hydroxyl radicals, the most damaging water radical produced by ionizing radiation, is of the order of 3 nm in the physiological environment of the cell [11], which is approximately equal to the 2.5-nm distance (0.83 mm in the ND gas phase) of the 50% ion extraction efficiency surface from the SV axis. One should note that the uncertainty of the parameters that are important for a biophysical model relating ion-cluster-size distributions to DNA damage yields, i.e., the probability of each registered ion to produce a specific damage (strand break or base damage), is relatively large and is mostly determined by the complexity of the chemical and biochemical processes following the physical stage. Additional work will be needed to demonstrate how sensitive biophysical models predicting the yield of DNA damage based on ND measurements are to parameters of the SV map.

In conclusion, both methods provide valuable information on the lateral fall-off of the ion extraction efficiency and the length of the high-efficiency region; these data are important for optimizing and interpreting the nanodosimetric measurements for the simulation of radiation effects in DNA segments [10]. The relatively good agreement between both approaches supports the use of the more practical calculational method for SV mapping in future works.

---

[3] Bernd Grosswendt, personal communication, 2005.

## Acknowledgements

This work was partially supported by the U.S. National Medical Technology Testbed Inc. (NMTB) under the U.S. Department of the Army Medical Research Acquisition Activity, Cooperative Agreement Number DAMD17-97-2-7016 and by the Minerva Foundation. The authors would like to thank William Preston, Ed.D for his assistance in preparing this manuscript. A.B. is the W.P. Reuther Professor of Research in the peaceful uses of Atomic Energy.

**Figure legends**

**Fig. 1.** Schematic cross section of the ion counting ND.

**Fig. 2.** Measured *D/K* ratios as a function of reduced field strength [5]. The dashed curves represent estimated maximum and minimum limits of the ratios.

**Fig. 3.** Simulated ion trajectories starting from a given point in the gas volume and equipotential lines above the aperture plate with 10 V and 1 V steps. The relative fraction of trajectories traversing the aperture corresponds to the ion extraction efficiency. The lateral spread due to ion diffusion and the focusing near the aperture are apparent.

**Fig. 4.** Correlation between the ion-drift time and the *y*-coordinate of the proton track measured with the STT. Each dot corresponds to a registered proton producing a registered ion. The straight lines represent the boundaries of the time-cut window applied to suppress ions generated by delta electrons; the dashed line corresponds to the level of the aperture plate.

**Fig. 5.** Measured (points) and Gaussian-fitted (line) mean number of ions per particle versus distance *x* from the SV axis at a vertical distance of *y* = 14.5 mm above the aperture.

**Fig. 6.** Integrated ion extraction efficiency, $\Sigma(x,y)$ evaluated with a Monte Carlo simulation of the scanning approach with a known SV map, with (open circle) and without (triangle) background and arrival-time corrections, as a function of distance *x* from the SV axis. It is compared with the true integrated efficiency (closed circles) of the known map at three different vertical *y* coordinates above the aperture.

**Fig. 7.** Color-shaded representation of the SV maps and their differences for 1.33mbar propane, stationary field extraction ($E_1$ = 60 V/cm) and 0.3 mm aperture plate thickness: (A) map obtained with the calculational approach; (B) map obtained with the scanning approach; (C) the difference between the two maps. Note the large difference in *x*- and *y*-scales.

**Fig. 8.** Color-shaded representation of the SV maps and their differences for 1.33mbar propane, pulsed-field extraction ($E_1$ = 20 V/cm plus 40 V/cm pulse) and 0.3 mm aperture plate thickness: (A) map obtained with the calculational approach; (B) map obtained with the scanning approach; (C) the difference between the two maps.

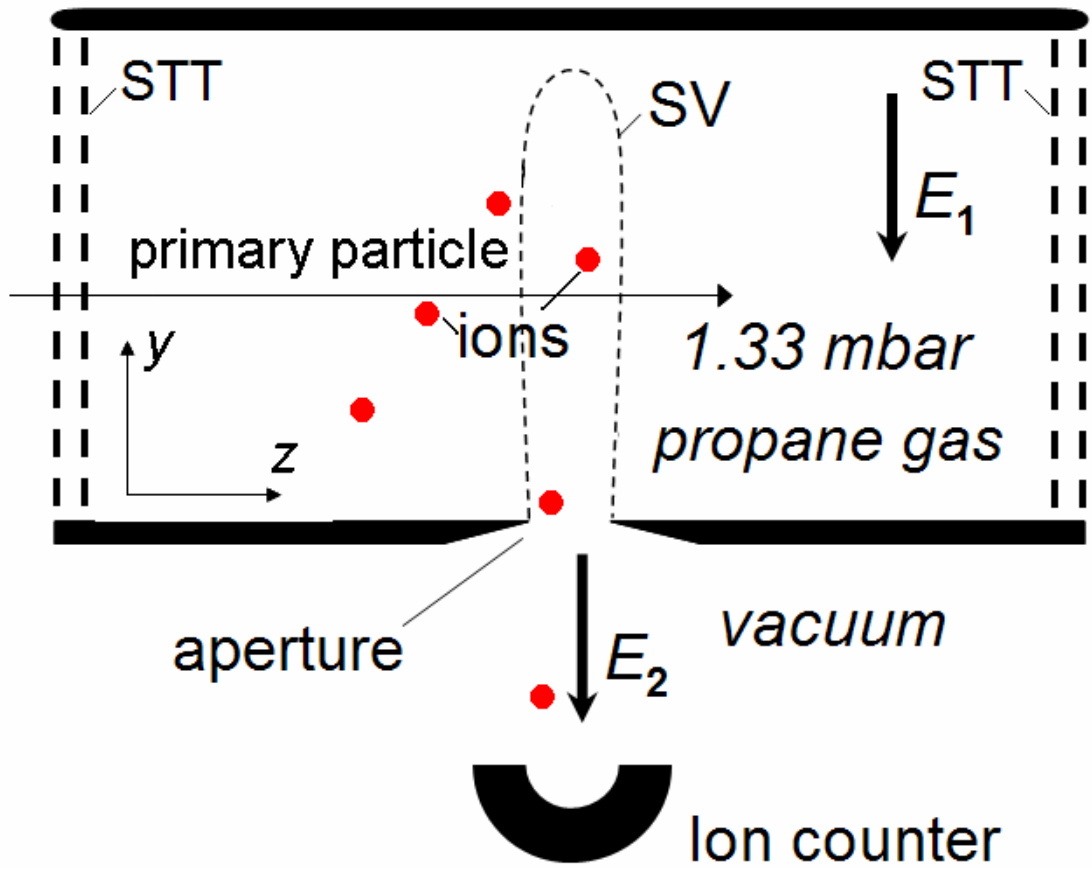

**Figure 1**

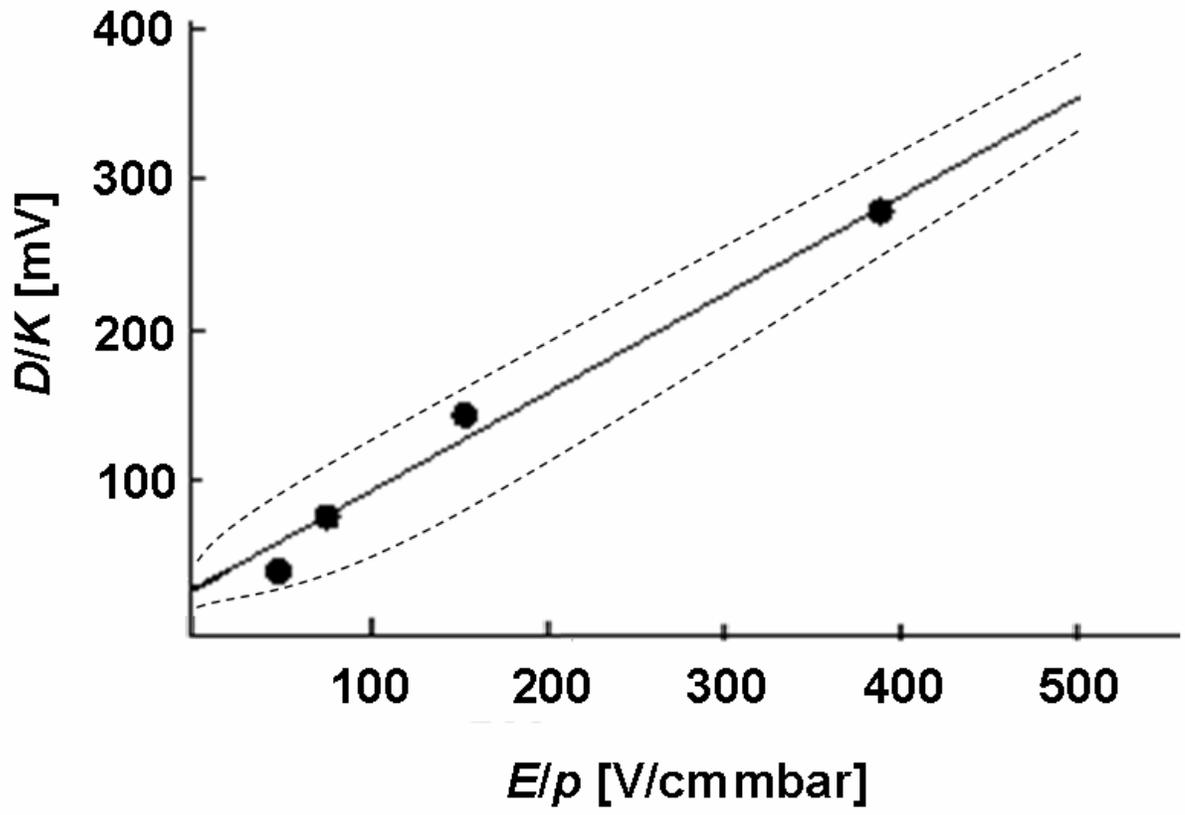

**Figure 2**

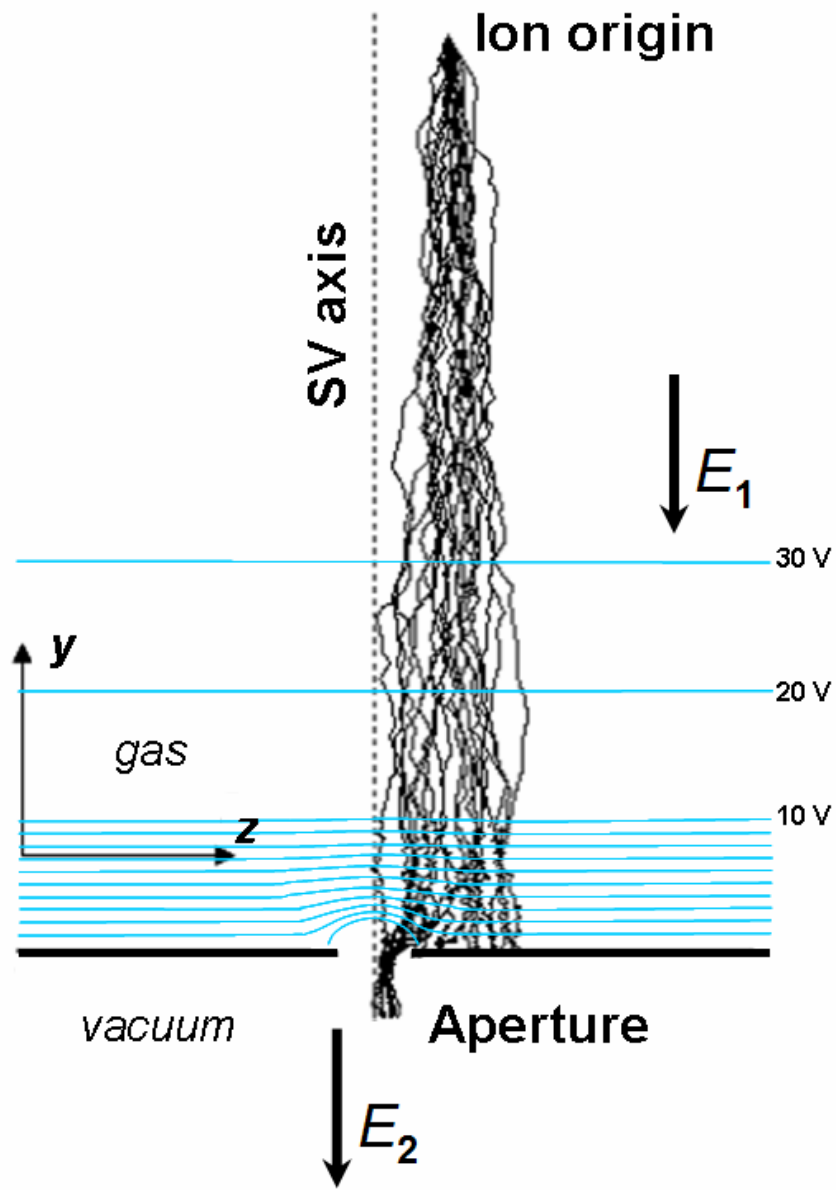

Figure 3

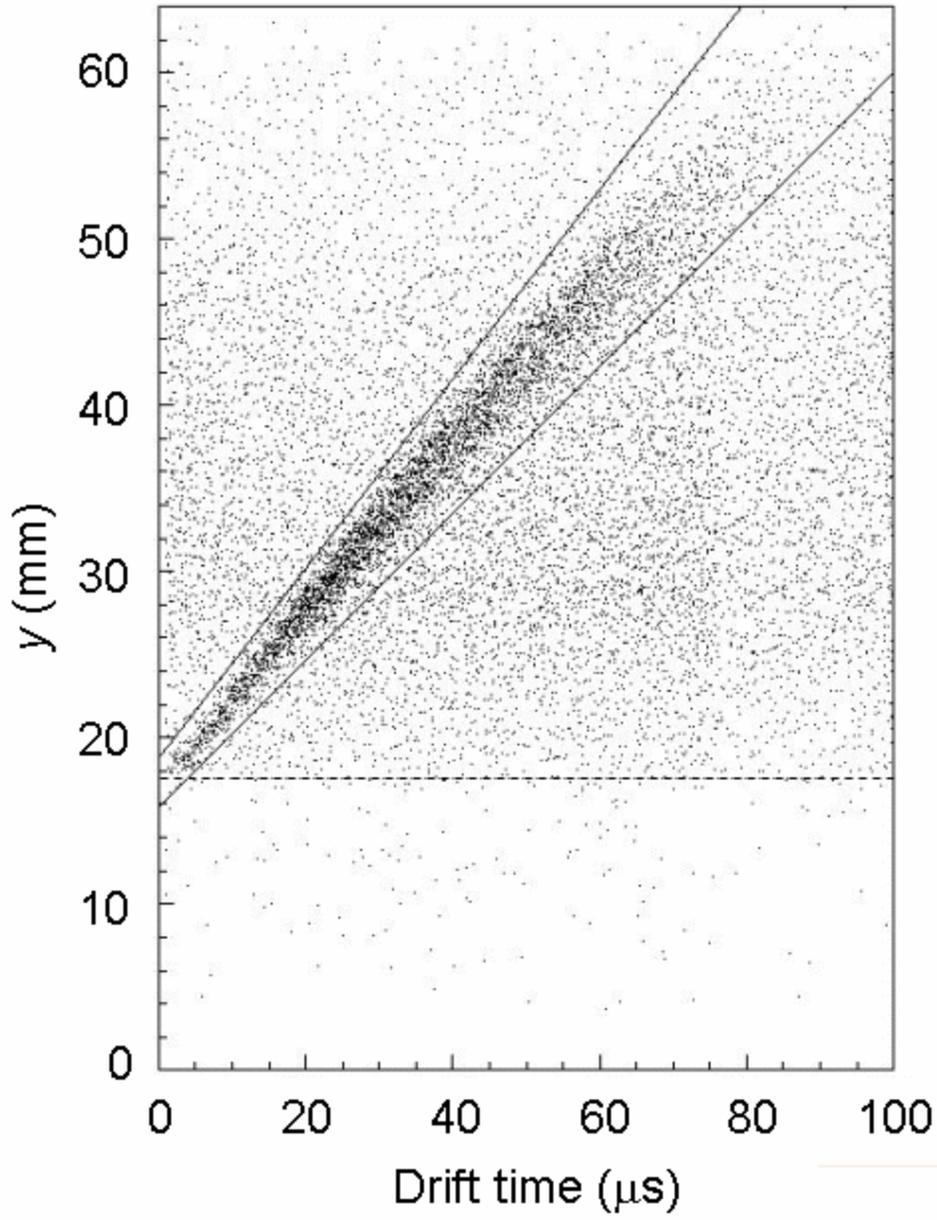

**Figure 4**

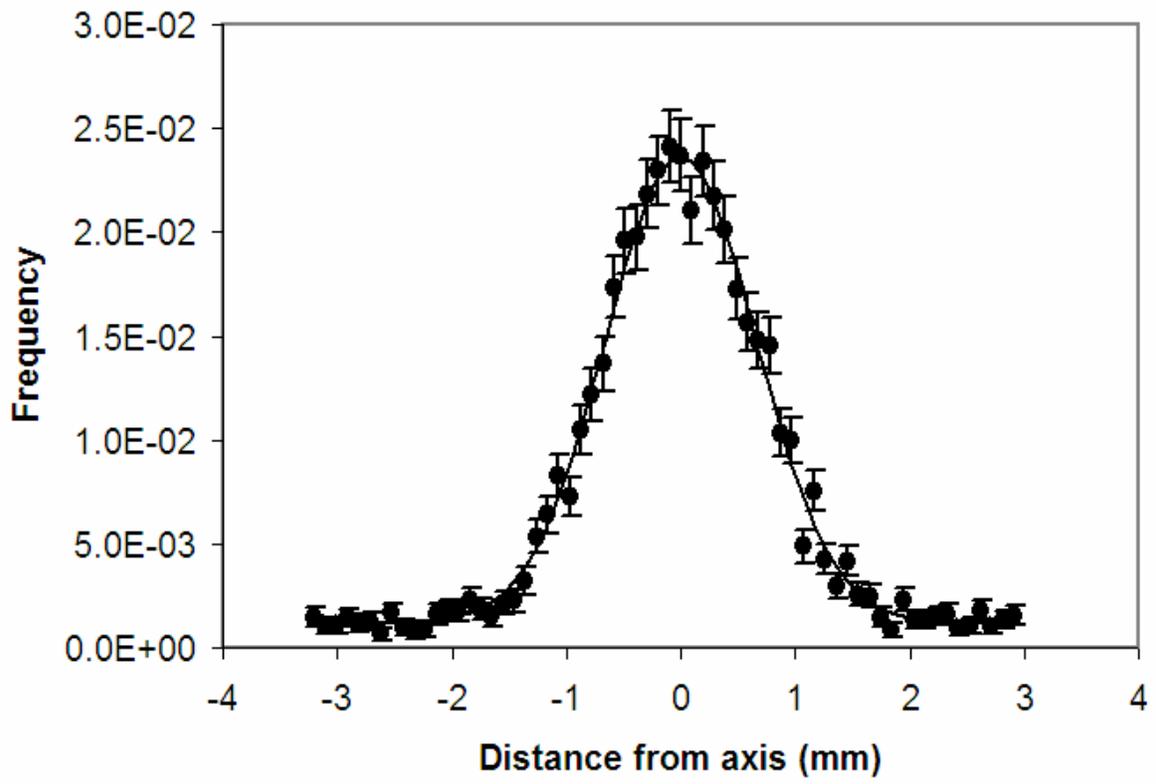

**Figure 5**

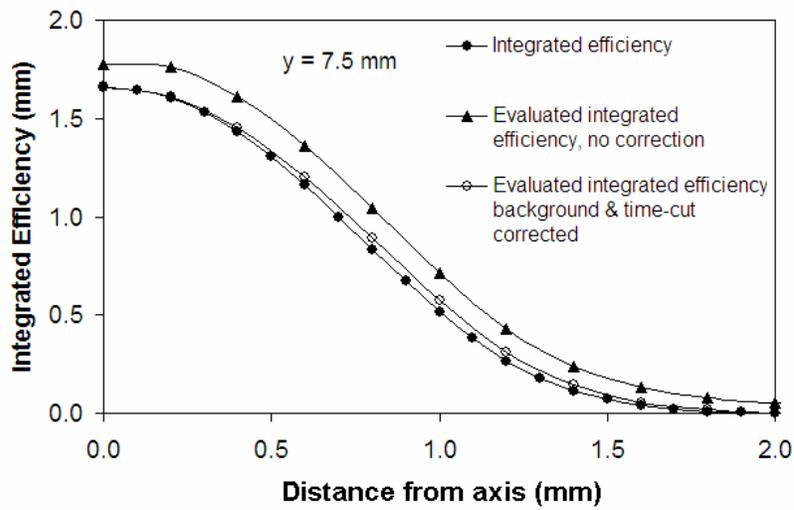
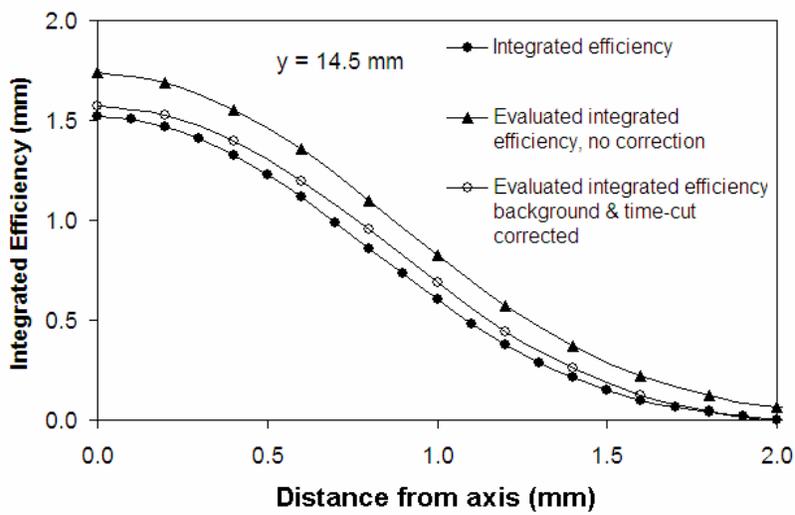
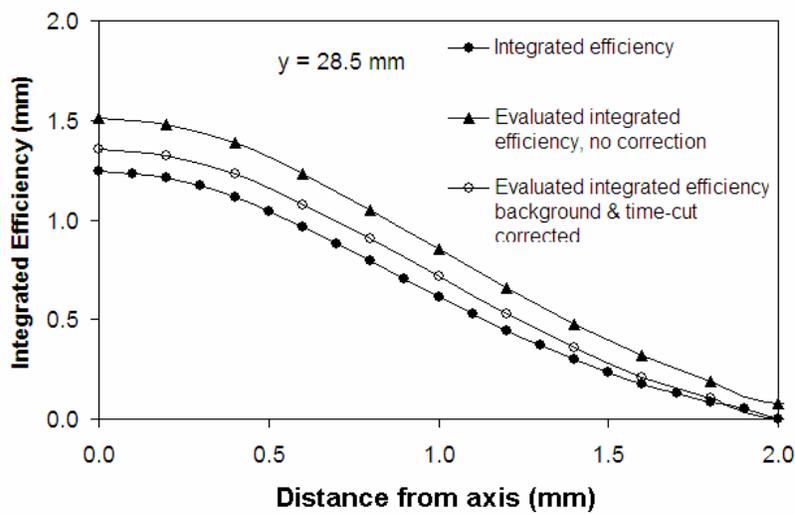

**Figure 6**

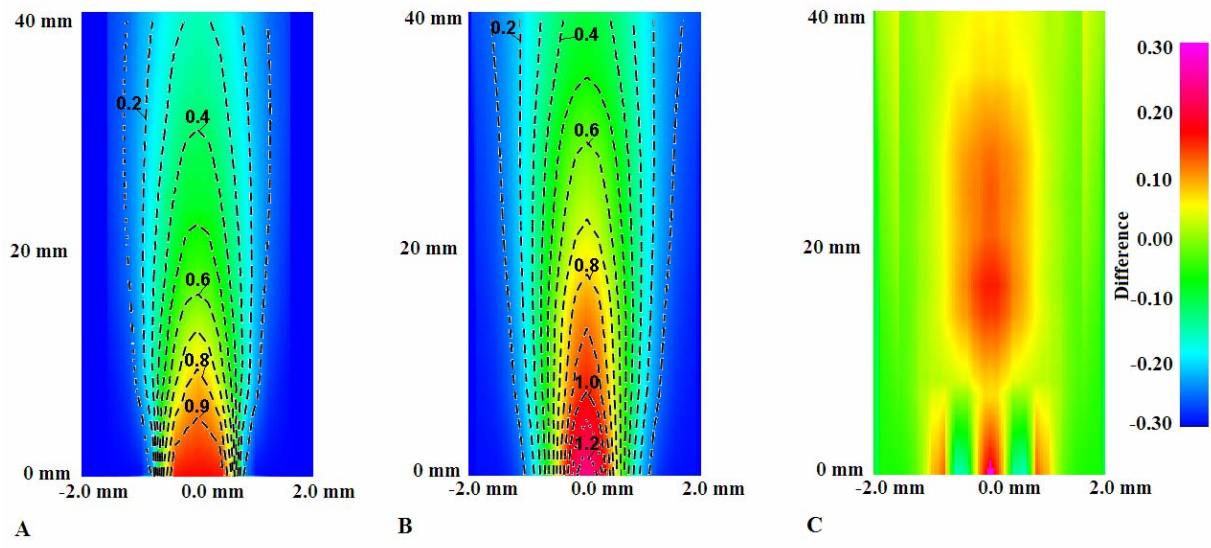

**Figure 7**

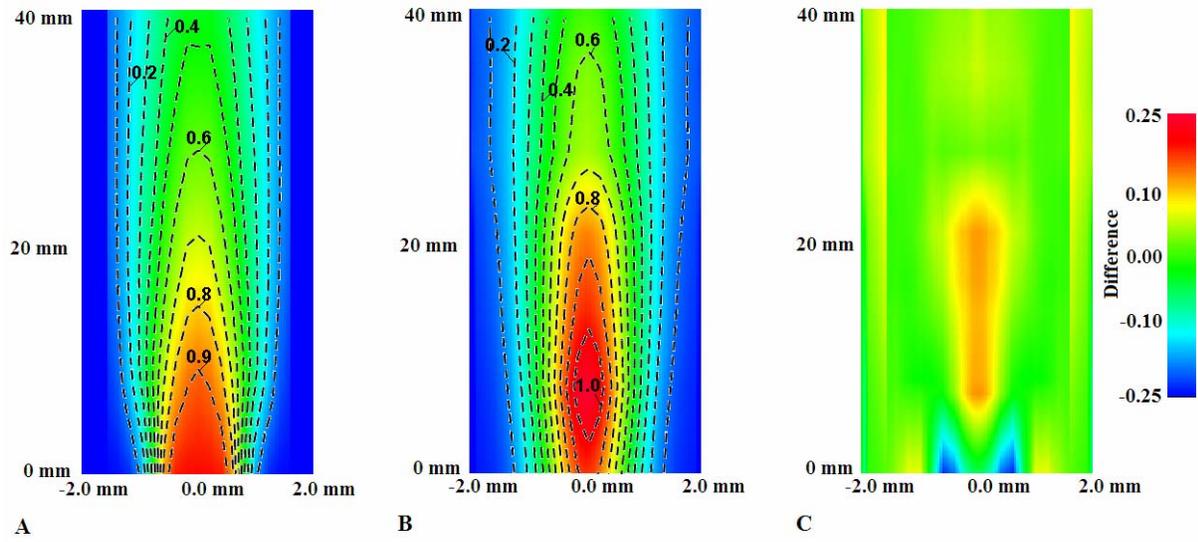

**Figure 8**